\documentclass{iopjournal}

\usepackage{subcaption} 
\usepackage{ragged2e}
\usepackage{amsmath}
\usepackage{amsfonts}

\begin{document}

\justifying
\articletype{Article type}

\title{Quantification of the parameter estimation error from Rotating Core Collapse supernovae}

\author{Claudia Moreno$^{1,*}$\orcid{0000-0002-0496-032X}, Javier~M.~Antelis$^2$\orcid{0000-0003-3377-0813} and Michele Zanolin$^{3}$\orcid{0000-0002-4044-4306)}}

\affil{$^1$Universidad de Guadalajara, Departamento de F\'isica CUCEI, Guadalajara, Jal., 44430, M\'exico}

\affil{$^2$Tecnologico de Monterrey, Escuela de Ingenieria y Ciencias, Monterrey, N.L., 64849, Mexico}

\affil{$^3$Embry-Riddle Aeronautical University, Prescott, AZ 86301, USA}

\affil{$^*$Universidad de Guadalajara, Departamento de F\'isica CUCEI, Guadalajara, Jal., 44430, M\'exico.}

\email{claudia.moreno@academico.udg.mx}


\begin{abstract}
\justifying{
In this paper, we perform parameter estimation with an analytical model to simulate the gravitational wave emission during the core-bounce phase of a rapidly rotating core collapse supernova progenitor. This approach enables us to estimate the parameter $\beta$, defined as the ratio of rotational kinetic energy to gravitational potential energy in core-collapse supernovae. To verify the reliability of both the analytical model and the inferred value of $\beta$, we use a numerical template bank constructed from Abylkairov’s gravitational waveform catalog and simulate O4 noise, characterized by the interferometer’s power spectral density.

An average fitting factor of 94\% over the interval 0.02 $< \beta <$ 0.14 shows that our analytical model reproduces the key characteristics of the core-bounce waveform with high accuracy, leading to only a 6\% reduction in the optimal signal-to-noise ratio. This provides a quantitative measure of how well the analytical model performs. Subsequently, we analyze the error in estimating $\beta$ using a Matched Filter method and compare it to the corresponding Cramér–Rao Lower Bound.  The results obtained by considering noise and waveforms at distances of 5, 10, and 50 kpc enable an assessment of how accurately the selected statistical model fits the observed data. From the asymptotic expansion of the variance, we derive a theoretical lower bound for the error that falls below $10^{-1}$ when the parameter $\beta$ decreases with distance.


}

\end{abstract}

\section{Introduction}

In recent years, following the first detection of gravitational waves due to the binary collision of black holes and neutron stars \cite{KAGRA:2021vkt} by the LIGO, Virgo and KAGRA (LVK) collaborations \cite{Aasi_2015, Acernese_2015, 10.1093/ptep/ptaa125}, supernova research  \cite{ Abdikamalov_2014, JANKA_2007, Kuroda_2016} has increased on several fronts, including numerical simulations, parameter estimations, and the development of multimessenger approaches \cite{Szczepa_czyk_2021, Lee_2015, corsi2024multimessengerastrophysicsblackholes}. 

Progenitor stars of supernovae are sometimes characterized by rapid rotation during their main-sequence stage \cite{Richers_2017, Ott_2012, Kotake_2006, Gilkis_2017, Couch_2014}. The typical rotational velocity at the equator is on the order of 200 km/s, representing a significant fraction of their breakup velocity, which can generate the core bounce. It is believed that hydrodynamic instabilities and rapid rotation in the cores of supernovae generate strong deviations from their spherical symmetry \cite{O_Connor_2018, Lella_2026}. Large-scale asymmetry in core rebound is induced by rotation and can convert some of the gravitational energy into gravitational waves.
In the domain of parameter estimation, several studies have been conducted to investigate the evolutionary and rotational dynamics of core-collapse supernova progenitor stars \cite{dimmelmeier2008gravitational, Pajkos_2021}. These studies primarily focus on understanding the alterations in rotational characteristics throughout the stellar lifecycle, culminating in the period immediately preceding the supernova event \cite{laura}. 

Numerical simulations aimed at predicting gravitational wave emission from core-collapse supernovae are computationally demanding and technically challenging to perform, and the resulting signals are predominantly stochastic in nature. In this work, we focus on gravitational waves generated by rapidly rotating core collapse supernovae \cite{Richers_2017, abylkairov2025evaluating}, particularly because they exhibit a deterministic core bounce component in the waveforms. This property enables the construction of an analytical model for estimating the physical parameter $\beta$ associated with the earliest stages of the core-collapse supernova process. To achieve this, we use an analytical model developed by Villegas et al. \cite{laura}. 
It is characterized by three physical parameters: the rotational ratio between kinetic and potential energy, $\beta = T/|V|$ \cite{Richers_2017, shibagaki2020new}; $\tau$, the time at which the bounce occurs; and $\alpha$, a parameter proportional to the amplitude of the third peak, which also incorporates the dispersion in peak amplitude induced by the equation of state across different rotational profiles. 
The analytical model was validated by numerical gravitational waves generated by Richers et al. in the catalog \cite{Richers_2017} and the references in it. It is shown there that the bounce of the gravitational wave is sensitive to the ratio of rotational to gravitational energy, and that the GW frequency of postbounce core oscillations shows a stronger dependence on the equation of state. The Ref. \cite{laura} aimed to establish, with the simplest possible analytical model, whether they can estimate $\beta$ with an error larger than a few percent. To estimate the value of $\hat \beta$ using the Matched Filter (MF) technique, they employ the Richers waveforms, which are injected into two cases: colored simulated Gaussian noise produced from the O3 Power Spectral Density (PSD) and real LIGO O3 noise from the Livingston detector.
Abdikairov et al. \cite{abylkairov2025evaluating} have recently generated a new catalog of gravitational-wave signals; these waveforms are employed in this study because their improved numerical errors enable more accurate parameter estimation. These waveforms exhibit a close morphology to the Richers et al. catalog. The waveforms presented examine the slow, moderate, and rapid rotation regimes, which are characterized by the occurrence of core bounce at supranuclear densities induced by centrifugal forces. They generate 452 General Relativity signals using four equations of state and 10.9 $M_{\odot}$ for the progenitor mass of the star. Using these simulations and following Villegas et al.'s analysis of the $\beta$ parameter in core-collapse supernovae, we employ the Abdikairov et al. simulation to estimate the value of $\beta$. We compare the relative error in $\beta$ derived from the theoretical variance model $\Delta \sigma = \sigma / \beta$ at first and second order with the uncertainties obtained via the Cramér–Rao Lower Bound and the MF approaches. The variance is evaluated as a function of the parameter $\beta$ for signals placed at distances of 5, 10, and 15 kpc, employing the power spectral density \cite{Moore_2014} of the generated data from the O4 \cite{T2000012}, Cosmic Explorer (CE) \cite{Srivastava_2019}, and Einstein Telescope (ET) \cite{Punturo_2010} detectors.

In this work, we begin our analysis in Sec. \ref{Sec:AsymptoticExpansions} where we review the theoretical lower bounds on parameter estimation. We present the Cramér-Rao Lower Bound and the asymptotic expansion approach for calculating the estimation error for the rotation parameter $\beta$ in Gaussian noise. 
In Sec. \ref{model} we introduce the phenomenological model designed to fit the core bounce component in terms of three exponential Gaussians. In Sec. \ref{Sec:Methodology}, We use the Fitting Factor ($FF$) to quantitatively assess the agreement between our analytical model and the corresponding numerical signal. We then employ a MF procedure to estimate the parameter $\hat{\beta}$ at radial distances of 5, 10, and 50 kpc.  Finally, in Sec. \ref{Sec:results}, we provide some comments and conclusions on this work.

\section{Error estimation based on the Cramér–Rao lower bound theorem} 
\label{Sec:AsymptoticExpansions}
\vspace{.1cm}
The Cramér-Rao lower bound theorem \cite{Zanolin_2010, Vitale_2010} states that for any unbiased parameter, its variance cannot be smaller than a specific bound, which is given by the inverse of the Fisher Information. 
In the context of gravitational waves, this represents the limit on how precisely we can infer a gravitational wave source’s properties (in our case, its rotation) from noisy data.

\vspace{.1cm}
Consider the $N$-dimensional observed dataset denoted as $\textbf{x} =\{x_{1}, \ldots ,x_{N}\}$, which depends on $P$ unknown parameters represented by the vector $\boldsymbol{\vartheta} = [\vartheta_1,\vartheta_2,\cdots,\vartheta_P ]^{T}$; where $T$ denotes the transposed vector.
In accordance with the Cramér-Rao lower bound theorem, the minimum achievable variance for each parameter is represented by $\mathrm{var}( \widehat{\vartheta_{i}} ) = \textbf{I}^{-1} (\boldsymbol{\vartheta})_{i j}$\footnote{Index runs $i,j=1,2,\cdots,P$.}, where $\textbf{I}( \boldsymbol{\vartheta} )$ constitutes the $P \times P$ Fisher Information Matrix defined by:
\begin{equation}\label{Equ:FIM}
         \textbf{I}_{i j} (\boldsymbol{\vartheta}) = - \operatorname{E} \left[ \dfrac{\partial^2 \ell (\textbf{x};\boldsymbol{\vartheta})}{\partial \vartheta_{i} \partial \vartheta_{j}} \right],
\end{equation}
where $\ell (\textbf{x};\boldsymbol{\vartheta}) = \ln\,{p(\textbf{x};\boldsymbol{\vartheta})}$ is the log-likelihood function of the unknown parameters given the observed data, and the operator $\operatorname{E}$ is the expected value of a random variable. Now, we need to obtain the estimated parameters $\widehat{ \boldsymbol{\vartheta} }$, where the observed data are random samples characterized by a probability density function $p(\textbf{x};\boldsymbol{\vartheta})$.

Given an observed dataset $\textbf{x}$, the Maximum Likelihood Estimate, denoted by $\widehat{\boldsymbol{\vartheta}}$, is obtained by selecting the parameter values that maximize the probability of the observed data. It is defined as follows:
\begin{equation}\label{Equ:MLE}
        \ell_{i} (\textbf{x};\boldsymbol{\vartheta}) = \dfrac{\partial \ell (\textbf{x};\boldsymbol{\vartheta})}{\partial \vartheta_{i} } \Bigm|_{ \boldsymbol{\vartheta} = \widehat{\boldsymbol{\vartheta}} } = 0,
\end{equation}
for regimes where the Maximum Likelihood Estimation is strongly unbiased and its covariance is much larger than the inverse of the Fisher Information Matrix. The second-order bias and second-order covariance allow us to control the reliability of the first-order approximation when evaluating the estimator. Additionally, it allows for the detection of divergent behavior in estimators \cite{Martynov2016}.

In asymptotic expansion, the total value of the bias estimator is given by 
\begin{equation}  
    \textbf{b}({\boldsymbol{\widehat \vartheta}_{i}} ) = \textbf{b}_{1}({\boldsymbol{\widehat \vartheta}_{i}} ) + \textbf{b}_{2}({\boldsymbol{\widehat \vartheta}_{i}} ) + \cdots \, ,
\end{equation}
where $\textbf{b}_{1}( {\boldsymbol{\widehat \vartheta}_{i}} )$ is the first order bias and $\textbf{b}_{2}({\boldsymbol{\widehat \vartheta}_{i}} )$ is the second order bias.


Analogously to the bias estimator, the covariance, denoted by $\sigma^2$, is expressed from the following expansion:
\begin{equation}
    \sigma^2_{ \widehat{\boldsymbol{\vartheta}}_{i} } = \sigma^2_{1}( \widehat{\boldsymbol{\vartheta}}_{i}) + \sigma^2_{2}( \widehat{\boldsymbol{\vartheta}}_{i}) + \cdots \, ,
\end{equation}
where $\sigma^2_{1}({\boldsymbol{\widehat \vartheta}}_{i})$ and $\sigma^2_{2}({\boldsymbol{\widehat \vartheta}}_{i})$ are the first and second order variances. 
In Ref. \cite{laura}, it was discussed how the MF is the MLE and how the bias and second order variance are negligible for the estimation of $\beta$.

\subsection{Additive noise in gravitational waves signals}
\vspace{.1cm}
Let us consider the data set obtained from a gravitational wave detector, represented as $x(t)$ and characterized as the sum of the signal model $h(t;\boldsymbol{\vartheta})$, a function dependent on $\boldsymbol\vartheta$, and the intrinsic noise present in the detector $n(t)$. Thus, the observed data can be expressed as
\begin{equation}\label{Equ:SignalPluNoise}
        x(t) = h(t;\boldsymbol{\vartheta}) + n(t).
\end{equation}
%
Next, we need to transition from the time domain to the frequency domain. Accordingly, we must express the Fisher Information Matrix in terms of the Fourier domain,
\begin{equation}
     \textbf{I} (\boldsymbol{\vartheta})_{i j} = \operatorname{E} \left[ \ell_{i} \ell_{j} \right] = \langle h_{i}(f), h_{j}(f) \rangle ,
     \label{FIM}
\end{equation}
being $h_{i}(f) = \partial h(f) / \partial \vartheta_i $ the partial derivative of $h(f) \equiv \int \,dt \exp^{-2\pi i f t }h(t;\boldsymbol{\vartheta})$.

By employing the asymptotic expansion method for estimating bias and covariance with respect to the single unknown parameter $\vartheta_{1} = \beta$, the expressions for first-order and second-order bias are represented as:
\begin{equation} \label{Eq:Bias_1}
    b_{1}( \beta ) = -\frac{1}{2} (I^{\beta\beta})^2 \langle h_{\beta\beta}, h_\beta \rangle \, ,
\end{equation}
\begin{eqnarray} \label{Eq:Bias_2}
    b_{2}( \beta ) &= - (I^{\beta\beta})^3 \left[\frac{1}{8} \langle h_{\beta\beta\beta\beta}, h_\beta\rangle + \frac{5}{4}\langle h_{\beta\beta\beta}, h_{\beta\beta}\rangle - \frac{3}{2}\langle h_{\beta\beta\beta}, h_{\beta}\rangle - I^{\beta\beta}\langle h_{\beta\beta\beta}, h_{\beta}\rangle \langle h_{\beta\beta}, h_{\beta}\rangle \right. \nonumber\\
    &\left. 
    - \frac{9}{2} I^{\beta\beta}\langle h_{\beta\beta}, h_{\beta\beta}\rangle\langle h_{\beta\beta}, h_{\beta} \rangle + \frac{9}{8}(I^{\beta\beta})^2\langle h_{\beta\beta}, h_{\beta}\rangle^3 \right].
\end{eqnarray}

The bias is the shift of the mean value of the
estimator with respect to the true value of the parameter. 
The components denoted as $h_{\beta\beta\beta\beta}$, $h_{\beta\beta\beta}$, $h_{\beta\beta}$, and $h_{\beta}$ represent the fourth, third, second, and first derivatives, respectively. The derivatives of the strain in the frequency domain, with respect to the parameter space, are defined as 
\begin{equation}
    h_{a,b,\cdots,P}(f) = \frac{\partial^P h(f)}{\partial \vartheta_a \partial \vartheta_b \cdots \partial \vartheta_P}.
\end{equation}

The first-order covariance associated with a single parameter is equivalent to the inverse of the Fisher information,
\begin{equation} \label{Ec:fisher_2}
 \sigma^2_1(\beta) = \frac{1}{I(\beta)_{\beta\beta}} \,,
\end{equation}
while the covariance squared is reduced to the expression
\begin{equation} \label{Ec:C2}
  \sigma^2_2(\beta) = \left(I^{\beta\beta}\right)^3 \left(\frac{7}{2} \langle h_{\beta\beta}, h_{\beta}\rangle^2 I^{\beta\beta} - \langle h_{\beta\beta\beta}, h_\beta \rangle \right) \,.
\end{equation}
Covariance analysis enables the quantification of uncertainties in parameter estimation and yields confidence intervals for the inferred true values of $\beta$ from a gravitational signal.

Eq. \eqref{FIM} can be written (for $i=j=\beta$) as 
\begin{equation}  \label{Eq:FI_one}
    I_{\beta\beta} = 4 \int^{f_{cut}}_{f_{low}} \frac{|h_\beta(f)|^2}{S_h(f)} df \,,
\end{equation}
where $S_h(f)$ is the power spectral density of the noise, is chosen so that all the frequencies where the integrands are positive and included.

%
%

\section{Core bounce analytical model}
\label{model}
\vspace{.1cm}

In rapidly rotating core-collapse supernovae, the gravitational-wave signal displays a deterministic structure during the first few milliseconds following core bounce. It features a sharp, generally dominant peak at the moment of bounce, produced by the abrupt deceleration of the inner core. Beyond the final peak, the signal contains stochastic and ring down features for which it is not clear if an analytical structure could be appropriate; consequently, we do not consider this region.

To characterize the performance of estimating the rotation parameter in a core-collapse supernova, we use the waveforms obtained by Abylkairov et al. \cite{abylkairov2025evaluating}. In their catalog, the simulations account for relativistic effects and generate 452 waveforms using 4 different equations of state and a star progenitor mass of 10.9 M$_\odot$. In this study, we focus on a subset of 100 waveforms that display distinct rotational characteristics, spanning slow, moderate, and rapid rotation regimes. All of the selected models are calculated within the framework of the conformally flat version of General Relativity and consistently include full relativistic effects.



The phenomenological model that we use for the bounce phase of the proto neutron star is defined by four parameters ($\beta$, $\tau$, $\eta$, $\alpha$) \cite{laura}; these parameters represent, in order, the rotational parameter, the arrival time, the Gaussian width, and the link between the signals and different equations of state. To model the core bounce with a simple, differentiable function, we use a sum of three Gaussian components, as 
\begin{equation} \label{Ec:signalmodel}
 h(t) = h_1(\beta)\,{\rm \exp}^{\left[-\frac{(t - \tau)^2}{2\eta^2} \right]} + h_2\,(\beta)\,{\rm \exp}^{\left[-\frac{(t - \tau_a)^2}{2\eta^2} \right]}
 + h_3(\alpha, \beta)\,{\rm \exp}^{\left[-\frac{(t - \tau_b)^2}{2\eta^2} \right]}\,,
\end{equation}
where the amplitudes of the peaks are denoted as 
\begin{eqnarray} \label{Ec:Amplitudes}
    h_1(\beta) &=& -13.2 +  2.89\times 10^3 \beta -1.31\times 10^4 \beta^2 \,, \nonumber \\
    h_2(\beta) &=& -1.03 - 5.52\times 10^3 \beta + 9.43\times 10^3 \beta^2 \, \nonumber \,, \nonumber \\
    h_3(\alpha, \beta) &=&  17.20 + \alpha \left(\frac{\beta}{0.06} \right)^2 \,. \nonumber
    \label{Ecu:h3}
\end{eqnarray}
{$h_1(\beta)$, $h_2(\beta)$, and $h_3(\alpha, \beta)$ are the amplitudes of the characteristic peaks in the gravitational signal}, $\eta = 0.2$ ms, and $\tau$ is the first peak arrival time. We define $\tau_a = \tau + 0.5 $ and $\tau_b= \tau + 1$ for $h_2(\beta) $ and $ h_3(\alpha,\beta)$. The first peak starts developing when $\tau$ is between -0.5 and -0.2 ms. The parameter $\alpha$ in h$_3(\alpha, \beta)$ is related to the difference in amplitude between the peaks for the equation of state.

The $\beta$ parameter is influenced by the proto neutron star’s rotational deformation and the gravitational restoring force of the core bounce; however, accurately characterizing this force requires combining intuitive rotational profiles with the supernova’s internal mass distribution.

\begin{figure}
 \centering
    \begin{subfigure}[b]{0.48\textwidth}
        \centering
        \includegraphics[width=0.99\textwidth]{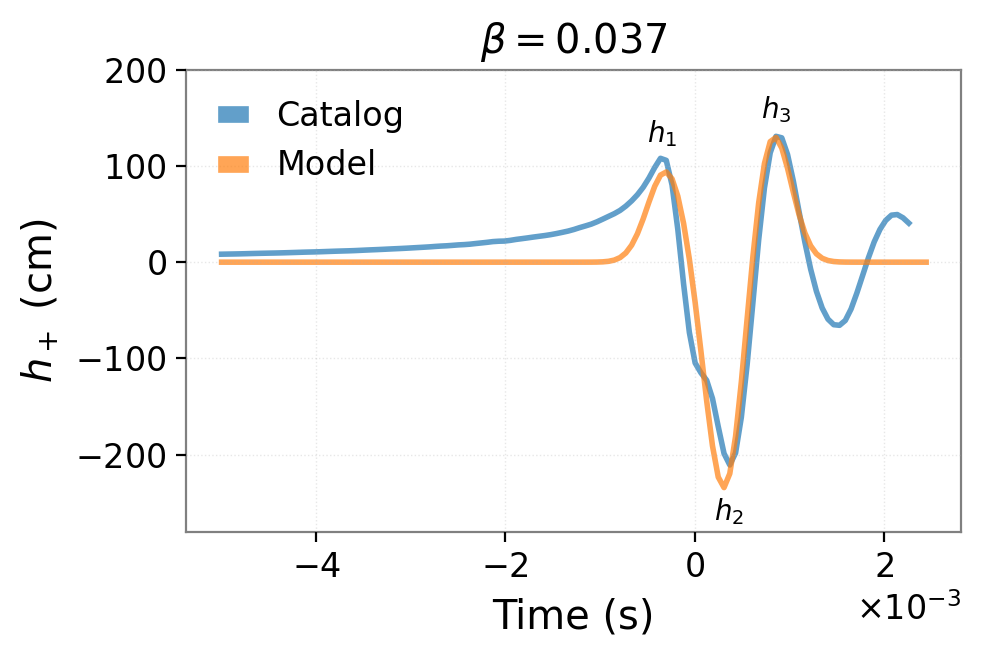}
        \caption{}
        \label{fig:comparison2}
    \end{subfigure}
    \hfill
    \begin{subfigure}[b]{0.48\textwidth}
        \centering
        \includegraphics[width=0.99\textwidth]{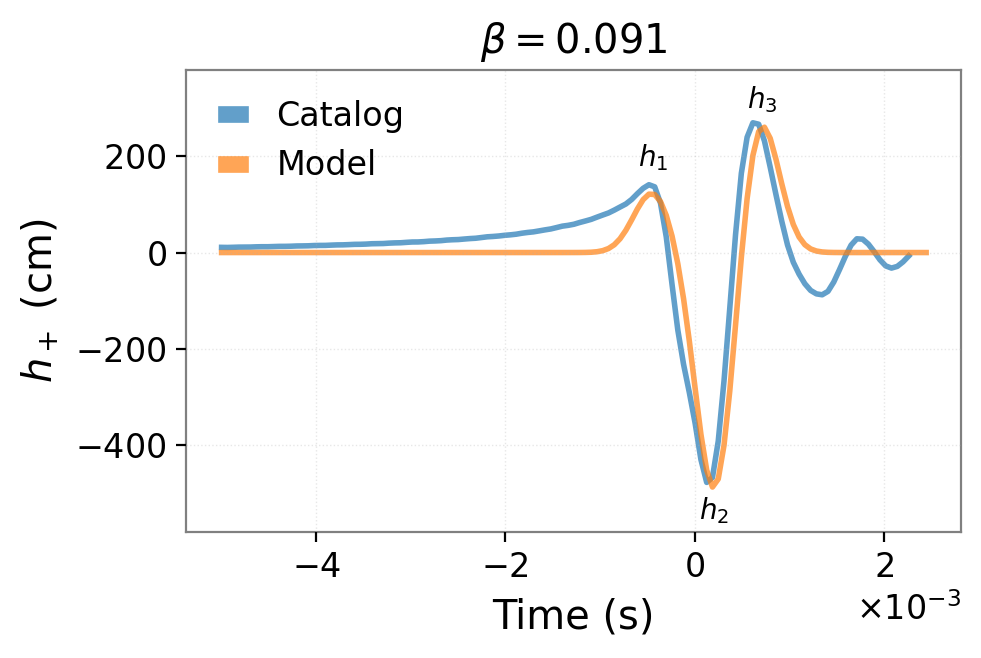}
        \caption{}
        \label{fig:comparison3}
    \end{subfigure}
\caption{
Examples of two gravitational waves signals from the Abylkairov catalog (blue) \cite{abylkairov2025evaluating}, along with their corresponding signal models (orange).
(a) Signal with a low rotation parameter value of $\beta=0.037$.
(b) Signal with a high rotation parameter value of $\beta=0.091$.
These two gravitational wave signals are employed in matched filter based parameter estimation to empirically determine the parameter estimation error.
}
\label{fig:ExampleSignals}
\end{figure}

In Fig. \ref{fig:ExampleSignals}, we show a comparison of one numerical waveform (blue) selected by Abykairov et al. from the catalog and a waveform produced by the analytical model (orange), Eq. \eqref{Ec:signalmodel}. In Fig. \ref{fig:comparison2}, we present the waveforms using the low rotation parameter $\beta=0.37$. The corresponding results for the high rotation parameter $\beta=0.91$ are displayed in Fig. \ref{fig:comparison3}. The gravitational waves show the characteristic three-peak structure of the core bounce phase. The agreement between the model and the numerical waveform is similar to that acquired in Ref. \cite{laura}. It is sufficient for an estimation of $\beta$ with an accuracy of a few tens of a percent, as discussed in the rest of the paper. 
We assess the parameter estimation uncertainties using matched filter based parameter estimation methods.

\begin{figure}
 \centering
    \begin{subfigure}[b]{0.48\textwidth}
        \centering
        \includegraphics[width=0.99\textwidth]{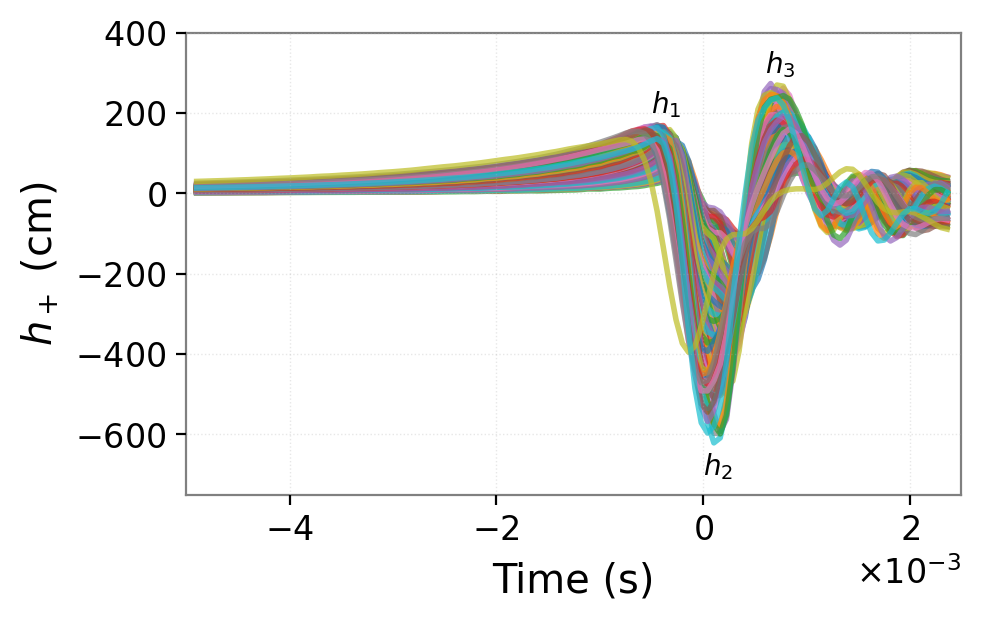}
        \caption{}
        \label{fig:waveforms}
    \end{subfigure}
    \hfill
    \begin{subfigure}[b]{0.48\textwidth}
        \centering
        \includegraphics[height=0.176\textheight]{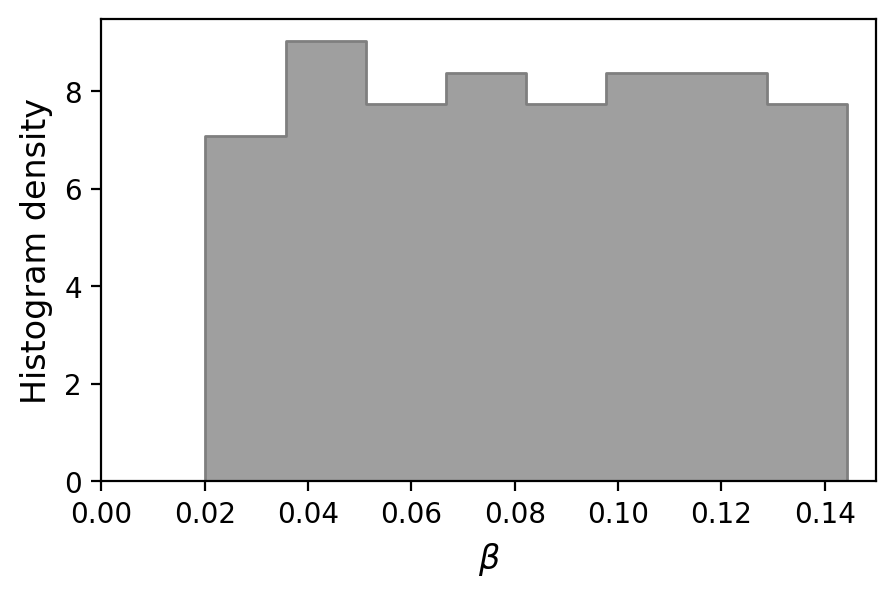}
        \caption{}
        \label{fig:histograma}
    \end{subfigure}
\caption{
(a) Set of 100 gravitational waves signals selected from the Abylkairov catalog for the $h_+$ polarization.
(b) Density histogram of the rotation parameter $\beta$ of the 100 selected gravitational waves signals. The plot shows a uniform distribution of $\beta$ in the interval $0.02$ - $0.14$.
}
\end{figure}

In Figure~\ref{fig:waveforms}, we present the temporal evolution of the plus polarization component of the gravitational wave strain, $h_+$, as a function of time. The considered parameter range is $\beta \in [0.02, 0.15]$, and time is measured in $[-0.005, 0.002]$ ms. We present a set of 100 signals chosen from the catalog of Abykairov et al. For the selected family of waveforms, we suppressed numerical artifacts by applying a low-pass filter, as described in Ref. \cite{laura}. These signals are obtained for different values of the parameter $\beta$, while the equation of state is kept fixed. As illustrated in the figure, most signals share a similar overall shape but differ in peak amplitudes. The changes in both amplitude and morphology among the various rotational profiles highlight how sensitive the gravitational wave signal is to the rotation parameter. For instance, the superimposed yellow signal does not display the third peak.  

In Fig. \ref{fig:histograma} we show a histogram that represents the distribution of the selected waveforms as a function of the parameter $\beta$. 
The signal density distributions are illustrated over the interval [0.02, 0.14], which encompasses the transition from slow rotation through moderate rotation (intermediate values of $\beta$) to rapid rotation. As shown, the histogram in Fig. \ref{fig:histograma} appears to be approximately uniform. 
This uniform sampling is appropriate for statistical analysis as it provides a balanced representation across the parameter space, allowing for an unbiased assessment of the $FF$ and parameter estimation accuracy across all rotational regimes.

\section{Results for estimating the rotational parameter}
\label{Sec:Methodology}
\vspace{.2cm}
In this section, we present the statistical methodology used to estimate the rotational parameter $\hat \beta$. 
Using Eq. \eqref{Ec:signalmodel}, we constructed an analytical template bank designed to identify the template that best matches a given gravitational wave numerical waveform from the Abylkairov et al. catalog and subsequently infer the value of the parameter $\beta$ for the waveform at distances of 5, 10, and 50 kpc.

\begin{figure}
 \centering
    \begin{subfigure}[b]{0.48\textwidth}
        \centering
        \includegraphics[width=0.99\textwidth]{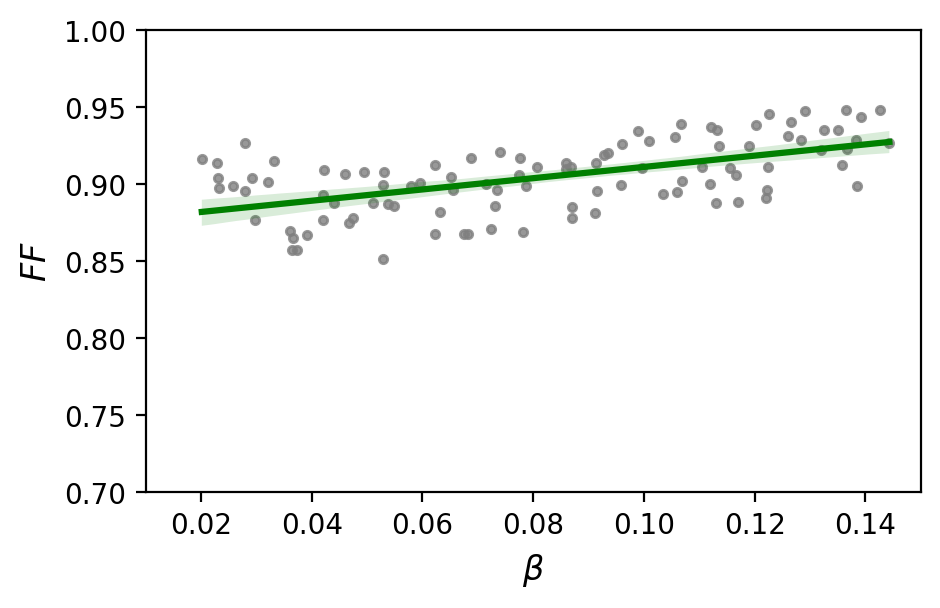}
        \caption{}
        \label{fig:sub1}
    \end{subfigure}
    \hfill
    \begin{subfigure}[b]{0.48\textwidth}
        \centering
        \includegraphics[width=0.99\textwidth]{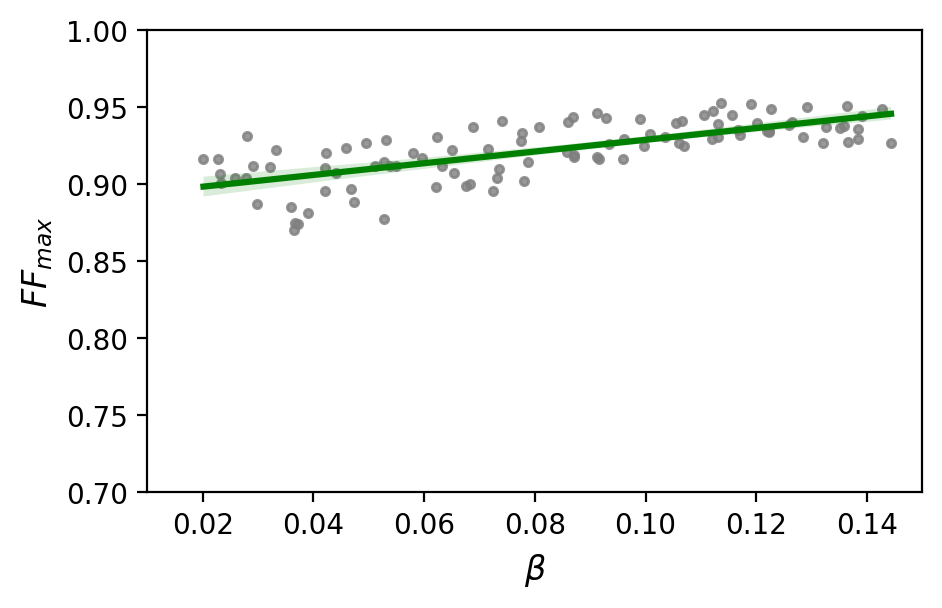}
        \caption{}
        \label{fig:sub2}
    \end{subfigure}
\caption{
%
(a) $FF$ value between each gravitational wave signal from the Abylkairov catalog and the corresponding gravitational wave signal model with the same value of the rotation parameter $\beta$.
(b) $FF_{\rm max}$ is obtained by tuning the waveforms in the template bank, determining, for each signal, the optimal combination of $\beta$ parameters that best matches the true waveform.}

\end{figure}

The first step in determining the rotational parameter is to apply the $FF$ method \cite{PhysRevD.52.605, Cho_2018, Sharma_2024}, which relies on the metric to quantify how well a modeled gravitational waveform agrees with its corresponding numerical signal. A high $FF$ implies that the template provides an accurate representation of the signal, whereas a low $FF$ indicates a poor match, potentially resulting in missed detections or biased parameter estimates. This quantity measures the fractional loss in signal-to-noise ratio incurred when employing our analytical template rather than the exact numerical waveform. A $FF$ of one corresponds to perfect agreement, implying that the template captures 100\% of the available signal power. In Fig. \ref{fig:sub1}, we examine how the factor varies between the simulated and numerical gravitational wave signals; as $\beta$ increases, the $FF$ becomes slightly larger. In Fig. \ref{fig:sub2}, we show the maximum $FF$ ($FF_{max}$) obtained by tuning the waveforms in the template bank, thereby determining, for each signal, the optimal combination of $\beta$ parameters that best matches the true waveform. We apply this procedure to each of the 100 chosen waveforms and determine the maximum $FF$ between the simulated signals and the template bank. In both cases, the $FF$ for the gravitational waves shows a modest linear increase as the rotation parameter $\beta$ grows. This behavior suggests that the analytical model reproduces the signals more accurately for systems with higher rotation rates. The slightly larger maximum $FF$ values relative to the $FF$ indicate that modest refinements to the estimated $\beta$ parameter can enhance the agreement between the templates and the numerical signals. An average $FF$ of 95\% over the range 0.02 $< \beta <$ 0.14 shows that our analytical model reproduces the main characteristics of the core-bounce waveform with high accuracy, leading to only a 5\% reduction in the optimal signal-to-noise ratio. This level of agreement is sufficient for both detection purposes and reliable parameter estimation and is consistent with the current level of uncertainty in numerical simulations of core-collapse supernovae.

%
%



\begin{figure}
 \centering
    \begin{subfigure}[b]{0.48\textwidth}
        \centering
        \includegraphics[width=0.70\textwidth]{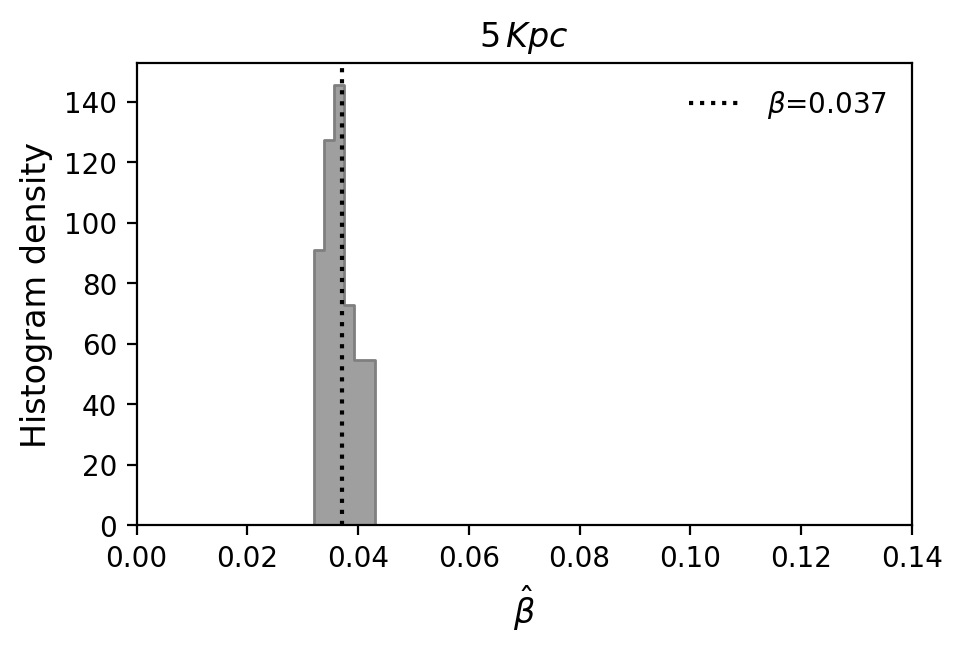}
    \end{subfigure}
    \hfill
    \begin{subfigure}[b]{0.48\textwidth}
        \centering
        \includegraphics[width=0.70\textwidth]{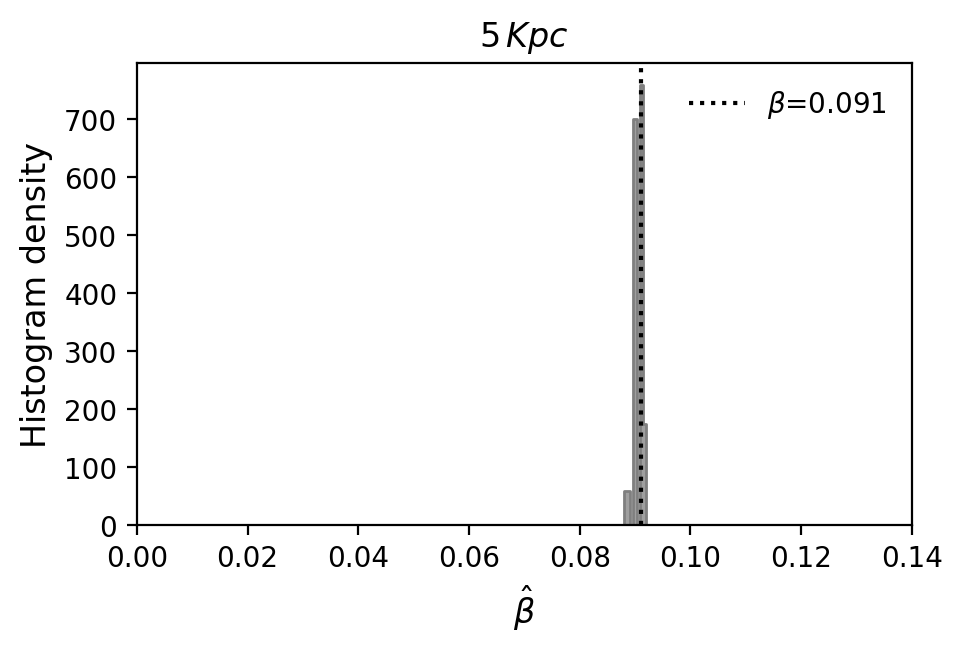}
    \end{subfigure}
    \\[1em]
    \begin{subfigure}[b]{0.48\textwidth}
        \centering
        \includegraphics[width=0.70\textwidth]{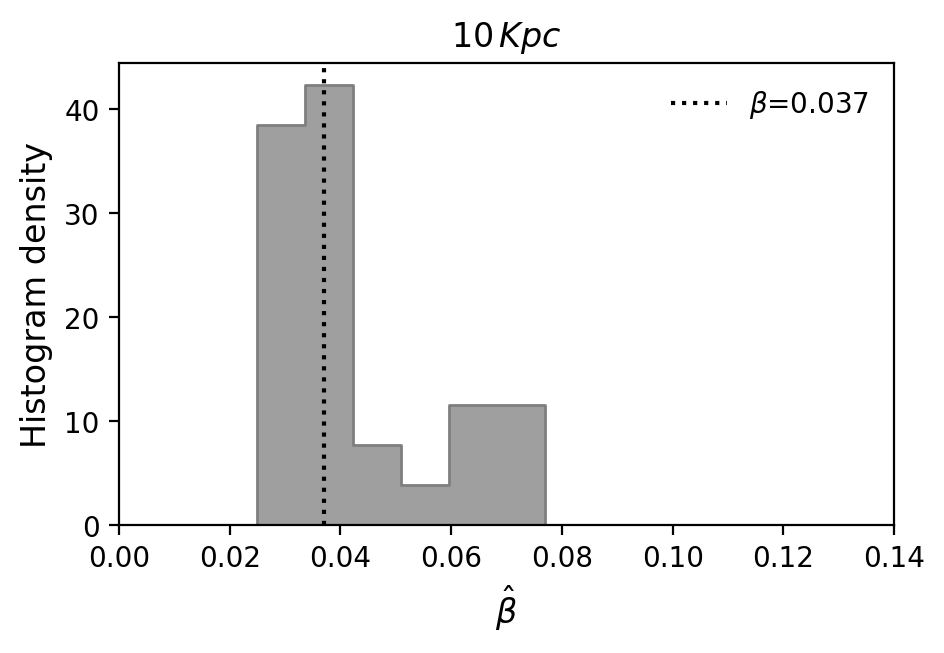}
    \end{subfigure}
    \hfill
    \begin{subfigure}[b]{0.48\textwidth}
        \centering
        \includegraphics[width=0.70\textwidth]{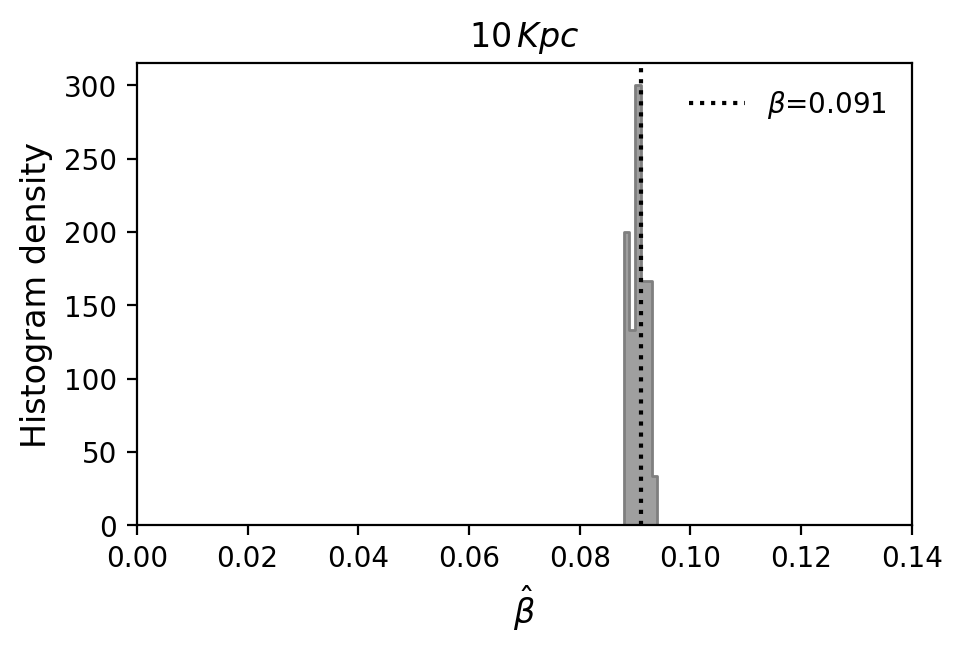}
    \end{subfigure}    
    \\[1em]
    \begin{subfigure}[b]{0.48\textwidth}
        \centering
        \includegraphics[width=0.70\textwidth]{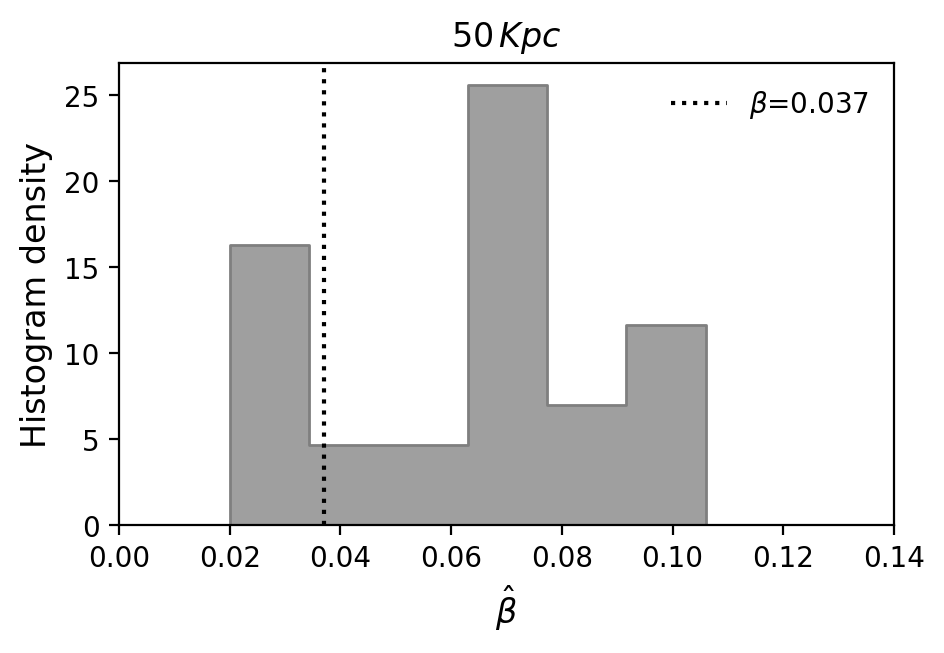}
        \caption{}
        \label{fig:histograms1}
    \end{subfigure}
    \hfill
    \begin{subfigure}[b]{0.48\textwidth}
        \centering
        \includegraphics[width=0.70\textwidth]{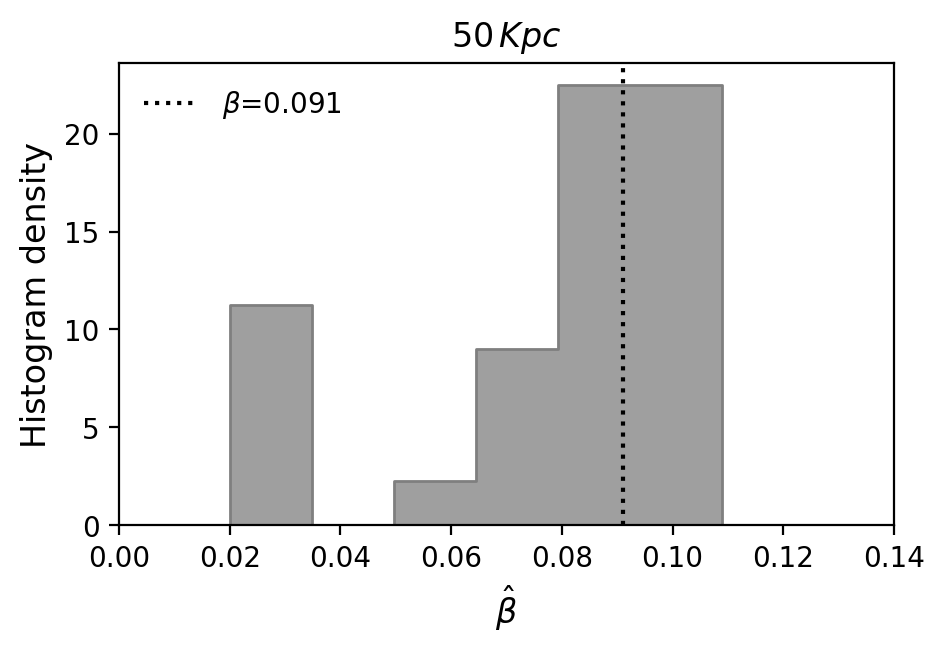}
        \caption{}
        \label{fig:histograms2}
    \end{subfigure}    
\caption{Histogram density of the estimated rotation parameter $\beta$ obtained with matched filter for the two gravitational waves signals shown in Figure \ref{fig:ExampleSignals} for the distances of 5, 10 and 50Kpc. 
(a) Estimation results for the gravitational wave signal from Abylkairov et al. with $\beta=0.037$; and (b) estimation with $\beta=0.091$.
%
%
In each histogram, the vertical black dotted lines represents the real $\beta$ value.
}
\label{fig:ResultadosFF}
\end{figure}

The selected signals, illustrated in Fig. \ref{fig:ExampleSignals}, were added to simulated Gaussian noise. To estimate the value of the $\hat \beta$ parameter for the injected gravitational wave signal corresponding to the chosen templates, we apply matched filtering to the signals $N = 1000$ times, each time with an independent realization of the noise, 
In Fig. \ref{fig:histograms1}, we present the histogram density of the inferred $\beta$ based on the selected waveform, which yields $\beta = 0.037$. This value lies in the slow-rotation regime, defined by $\beta < 0.08$. The second case, shown in Fig. \ref{fig:histograms2}, corresponds to the waveform with $\beta = 0.091$. This value lies in the rapid rotation regime, which is defined by $\beta > 0.08$. 
The histograms display the distribution of the estimated values $\hat{\beta}$ obtained from the simulations, along with the true value $\beta$ indicated by a vertical black dotted line. The histogram indicates that the matched filter estimator is essentially unbiased, as the mean of the estimates is concentrated near the true parameter values. Furthermore, the results show that the variance of the estimates increases systematically with distance, reflecting a weakening of the signal. At 5 kpc, the inferred values lie tightly around the $\beta$ value, and at 50 kpc, the distributions broaden considerably, indicating that parameter estimates become less precise at larger distances. The relatively narrower distributions observed for the rapid rotation configuration $\beta = 0.091$ suggest that signals with larger rotational parameters can enable more precise and strongly constrained parameter estimation.

\begin{figure}
 \centering
    \begin{subfigure}[b]{0.48\textwidth}
        \centering
        \includegraphics[width=0.99\textwidth]{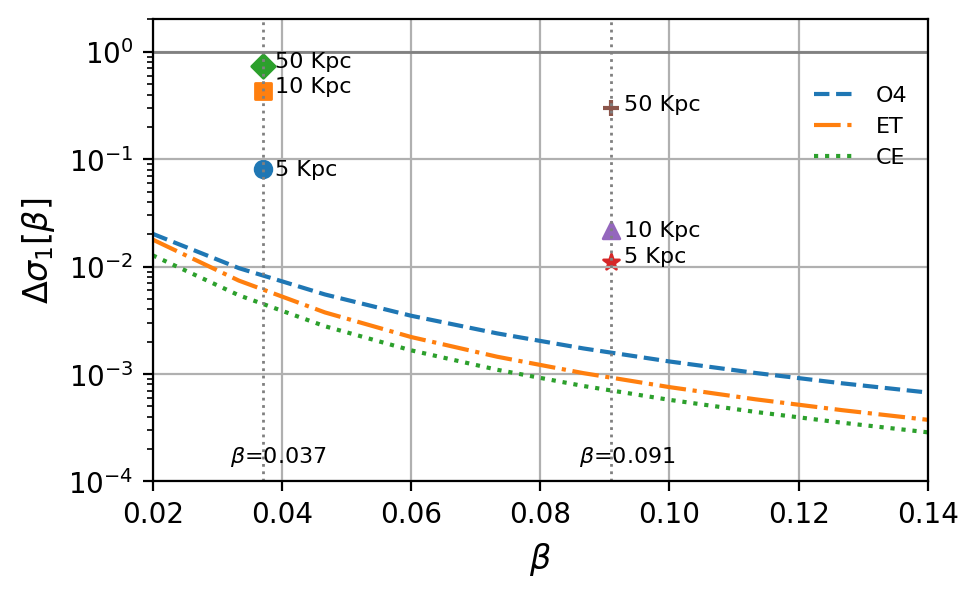}
        \caption{}
        \label{fig:comparison}
    \end{subfigure}
    \hfill
    \begin{subfigure}[b]{0.48\textwidth}
        \centering
        \includegraphics[width=0.99\textwidth]{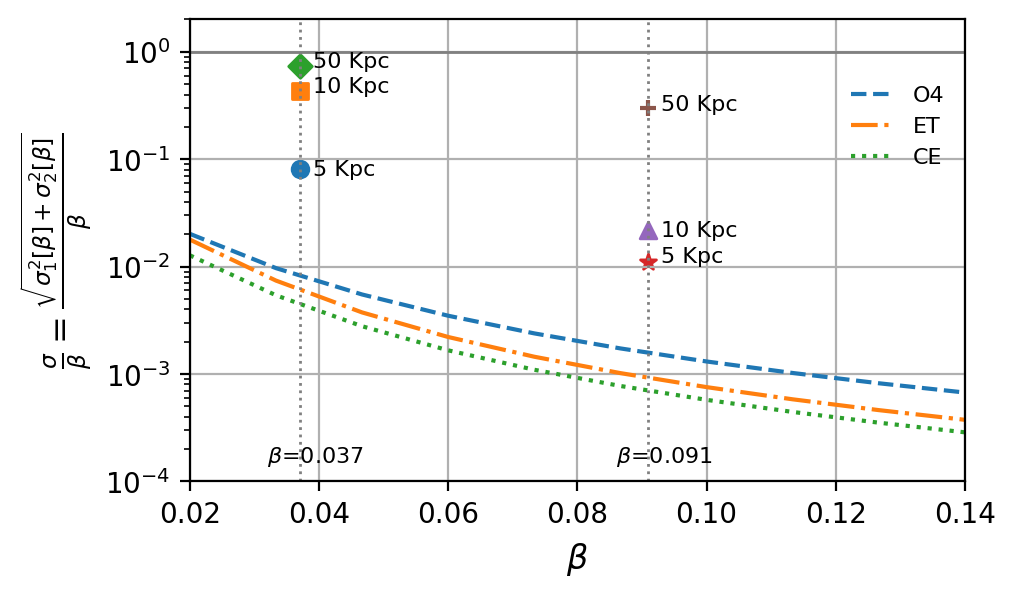}
        \caption{}
        \label{fig:comparison1}
    \end{subfigure}
\caption{
Relative error based on the variance, as a function of the parameter $\beta$ for signals at a distance of 5, 10 and 15 kpc for O4, ET and CE detector. The horizontal gray line indicates a relative error equal to one. (a) correspond to relative error of the first order variance $\Delta \sigma_1$. The dots correspond to variance of signal $\beta=0.037$ and $\beta=0.091$ obtained from MF with Gaussian noise. (b) We derive the second-order total approximation of the estimation error in order to quantify the accuracy of the results produced by our phenomenological model.}
\label{fig:Examplesigma}
\end{figure}

In Fig. \ref{fig:comparison}, we show the variance with respect to the parameter $\beta$ as $\Delta \sigma_1 = \sigma_1 /\beta$ and $\Delta \sigma_2 = \sigma_2 /\beta$, using the power spectral density of O4, CE, and ET detectors at 5, 10, and 50 kpc. The horizontal gray line indicates the theoretical minimum. The relative error of the second order covariance is smaller here than that of the first order, making it less relevant and therefore not plotted. We add dots in the images that correspond to the relative error of the variance of the signal at $\beta=0.037$ and $\beta=0.091$
obtained from MF with Gaussian noise. 
In Fig. \ref{fig:comparison1}, we obtain the estimation error,
\begin{equation}
    \frac{\sigma}{\beta} = \frac{\sqrt{\sigma^2_1 + \sigma^2_2}}{\beta},
\end{equation}
to measure the accuracy of the results obtained by our phenomenological model. Comparing Fig. \ref{fig:comparison} and \ref{fig:comparison1}, we see that the contribution of the second order variance is negligible for the estimation error, as discussed for Fig. \ref{fig:Examplesigma}. 
The figure also shows that, due to its improved sensitivity, the relative error is smaller for the CE detector than for O4.

%
%

\section{Discussions and Conclusions}
\label{Sec:results}
\vspace{.1cm}
In this study, we concentrated on extracting the rotational parameter $\beta$ from gravitational wave signals generated during the core-bounce phase of slow and rapidly rotating core-collapse supernovae from the Abylkairov et al. catalog. We employed the Cramér–Rao Lower Bound theorem to derive theoretical limits on the accuracy of analytical parameter estimation. An estimator that meets the theoretical minimum is not guaranteed to exist. In this case, the residual differences between the templates and the numerical model might have a role as well. This analysis offers insights into how estimation errors behave in different regions of the parameter space and identifies fundamental bounds on the attainable precision imposed by the properties of detector noise.

We adopt a phenomenological analytical model, expressed as the sum of three exponential Gaussians, to describe the morphology of core-bounce gravitational waves. To test this analytical framework, we draw on a set of 100 numerical waveforms from the Abylkairov et al. catalog, covering rotational regimes from slow ($\beta <$ 0.08) to rapid rotation ($\beta >$ 0.08). The fitting factor analysis revealed an average agreement of approximately 95\% between our analytical templates and the numerical simulations, demonstrating that our model successfully captures the essential features of the core-bounce signal across the full range of rotation parameters considered. Finally, employing matched filtering methods in the presence of Gaussian noise, we carried out parameter estimation of the rotation parameter $\beta$ for source distances of 5, 10, and 50 kpc. For each waveform, we generated N = 1000 independent noise realizations to obtain reliable statistical distributions for the estimated parameter $\hat \beta$. Our results demonstrate that the matched filter estimator is approximately unbiased, with mean estimates clustering near the true parameter values. As expected, the estimation precision degrades systematically with increasing distance due to the corresponding decrease in the signal to noise ratio. Notably, waveforms with higher $\beta$ values exhibit improved parameter estimation accuracy. 
From matched filter analysis, we conclude that the relative error obtained for generated noise from O4, CE, and ET data at distances of 5, 10, and 15 kpc is on the order of $10^{-2}$. A relative error smaller than one indicates the capability to resolve that beta is different from zero and that rotation is present in the progenitor. It is worth noting that the O4 data were obtained using the power spectral density of the LIGO noise, which is very similar to the actual O4a data. Therefore, we expect that our results will not differ significantly in future studies using real O4 data.

Future studies could build on this analysis by including additional waveform catalogs spanning a broader range of progenitor masses to assess model robustness; creating joint parameter estimation frameworks that simultaneously constrain rotation, the equation of state, and progenitor mass; and examining how waveform uncertainties and systematic modeling errors influence parameter estimation.


\ack{
This research has made use of data or software obtained from the Gravitational Wave Open Science Center (gwosc.org), a service of the LIGO Scientific Collaboration, the Virgo Collaboration, and KAGRA. This material is based upon work supported by NSF's LIGO Laboratory which is a major facility fully funded by the National Science Foundation, as well as the Science and Technology Facilities Council (STFC) of the United Kingdom, the Max-Planck-Society (MPS), and the State of Niedersachsen/Germany for support of the construction of Advanced LIGO and construction and operation of the GEO600 detector. Additional support for Advanced LIGO was provided by the Australian Research Council. Virgo is funded, through the European Gravitational Observatory (EGO), by the French Centre National de Recherche Scientifique (CNRS), the Italian Istituto Nazionale di Fisica Nucleare (INFN) and the Dutch Nikhef, with contributions by institutions from Belgium, Germany, Greece, Hungary, Ireland, Japan, Monaco, Poland, Portugal, Spain. KAGRA is supported by Ministry of Education, Culture, Sports, Science and Technology (MEXT), Japan Society for the Promotion of Science (JSPS) in Japan; National Research Foundation (NRF) and Ministry of Science and ICT (MSIT) in Korea; Academia Sinica (AS) and National Science and Technology Council (NSTC) in Taiwan. C.M. wants to thank SNII-SECIHTI. }

\funding{M.Z. is supported by the National Science Foundation Gravitational Physics Experimental and Data Analysis Program through awards PHY-2110555 and PHY-2405227.}

\roles{Conceptualization, J.M.A, C.M. and M.Z.; methodology, J.M.A, C.M. and M.Z.; formal
analysis, J.M.A, C.M. and M.Z.; investigation, J.M.A, C.M. and M.Z.; writing - original draft
preparation, J.M.A, C.M. and M.Z.; writing - review and editing, J.M.A, C.M. and M.Z. All authors have read and agreed to the published version of the manuscript.}

\data{No new data was obtained.}

\bibliographystyle{vancouver}
\bibliography{bibliography}

@article{abylkairov2025evaluating,
doi = {10.1088/2632-2153/ada33a},
url = {https://doi.org/10.1088/2632-2153/ada33a},
year = {2025},
month = {jan},
publisher = {IOP Publishing},
volume = {5},
number = {4},
pages = {045077},
author = {Abylkairov, Y Sultan and Edwards, Matthew C and Orel, Daniil and Mitra, Ayan and Shukirgaliyev, Bekdaulet and Abdikamalov, Ernazar},
title = {Evaluating machine learning models for supernova gravitational wave signal classification},
journal = {Machine Learning: Science and Technology}
}

@article{Acernese_2015,
    author = "Aasi, J. and others",
    collaboration = "LIGO Scientific",
    title = "{Advanced LIGO}",
    eprint = "1411.4547",
    archivePrefix = "arXiv",
    primaryClass = "gr-qc",
    doi = "10.1088/0264-9381/32/7/074001",
    journal = "Class. Quant. Grav.",
    volume = "32",
    pages = "074001",
    year = "2015"
}

@article{PhysRevD.52.605,
  title = {Search templates for gravitational waves from precessing, inspiraling binaries},
  author = {Apostolatos, Theocharis A.},
  journal = {Phys. Rev. D},
  volume = {52},
  issue = {2},
  pages = {605--620},
  numpages = {0},
  year = {1995},
  month = {Jul},
  publisher = {American Physical Society},
  doi = {10.1103/PhysRevD.52.605},
  url = {https://link.aps.org/doi/10.1103/PhysRevD.52.605}
}

@article{Abdikamalov_2014,
   title={Measuring the angular momentum distribution in core-collapse supernova progenitors with gravitational waves},
   volume={90},
   ISSN={1550-2368},
   url={http://dx.doi.org/10.1103/PhysRevD.90.044001},
   DOI={10.1103/physrevd.90.044001},
   number={4},
   journal={Physical Review D},
   publisher={American Physical Society (APS)},
   author={Abdikamalov, Ernazar and Gossan, Sarah and DeMaio, Alexandra M. and Ott, Christian D.},
   year={2014},
   month={Aug}
}

@article{Aasi_2015,
    author = "Aasi, J. and others",
    collaboration = "LIGO Scientific",
    title = "{Advanced LIGO}",
    eprint = "1411.4547",
    archivePrefix = "arXiv",
    primaryClass = "gr-qc",
    doi = "10.1088/0264-9381/32/7/074001",
    journal = "Class. Quant. Grav.",
    volume = "32",
    pages = "074001",
    year = "2015"
}

@article{10.1093/ptep/ptaa125,
    author = {Akutsu, T and Ando, M and Arai, K and Arai, Y and Araki, S and Araya, A and Aritomi, N and Aso, Y and Bae, S and Bae, Y and Baiotti, L and Bajpai, R and Barton, M A and Cannon, K and Capocasa, E and Chan, M and Chen, C and Chen, K and Chen, Y and Chu, H and Chu, Y -K and Eguchi, S and Enomoto, Y and Flaminio, R and Fujii, Y and Fukunaga, M and Fukushima, M and Ge, G and Hagiwara, A and Haino, S and Hasegawa, K and Hayakawa, H and Hayama, K and Himemoto, Y and Hiranuma, Y and Hirata, N and Hirose, E and Hong, Z and Hsieh, B H and Huang, C -Z and Huang, P and Huang, Y and Ikenoue, B and Imam, S and Inayoshi, K and Inoue, Y and Ioka, K and Itoh, Y and Izumi, K and Jung, K and Jung, P and Kajita, T and Kamiizumi, M and Kanda, N and Kang, G and Kawaguchi, K and Kawai, N and Kawasaki, T and Kim, C and Kim, J C and Kim, W S and Kim, Y -M and Kimura, N and Kita, N and Kitazawa, H and Kojima, Y and Kokeyama, K and Komori, K and Kong, A K H and Kotake, K and Kozakai, C and Kozu, R and Kumar, R and Kume, J and Kuo, C and Kuo, H -S and Kuroyanagi, S and Kusayanagi, K and Kwak, K and Lee, H K and Lee, H W and Lee, R and Leonardi, M and Lin, L C -C and Lin, C -Y and Lin, F -L and Liu, G C and Luo, L -W and Marchio, M and Michimura, Y and Mio, N and Miyakawa, O and Miyamoto, A and Miyazaki, Y and Miyo, K and Miyoki, S and Morisaki, S and Moriwaki, Y and Nagano, K and Nagano, S and Nakamura, K and Nakano, H and Nakano, M and Nakashima, R and Narikawa, T and Negishi, R and Ni, W -T and Nishizawa, A and Obuchi, Y and Ogaki, W and Oh, J J and Oh, S H and Ohashi, M and Ohishi, N and Ohkawa, M and Okutomi, K and Oohara, K and Ooi, C P and Oshino, S and Pan, K and Pang, H and Park, J and Arellano, F E Peña and Pinto, I and Sago, N and Saito, S and Saito, Y and Sakai, K and Sakai, Y and Sakuno, Y and Sato, S and Sato, T and Sawada, T and Sekiguchi, T and Sekiguchi, Y and Shibagaki, S and Shimizu, R and Shimoda, T and Shimode, K and Shinkai, H and Shishido, T and Shoda, A and Somiya, K and Son, E J and Sotani, H and Sugimoto, R and Suzuki, T and Suzuki, T and Tagoshi, H and Takahashi, H and Takahashi, R and Takamori, A and Takano, S and Takeda, H and Takeda, M and Tanaka, H and Tanaka, K and Tanaka, K and Tanaka, T and Tanaka, T and Tanioka, S and Tapia San Martin, E N and Telada, S and Tomaru, T and Tomigami, Y and Tomura, T and Travasso, F and Trozzo, L and Tsang, T and Tsubono, K and Tsuchida, S and Tsuzuki, T and Tuyenbayev, D and Uchikata, N and Uchiyama, T and Ueda, A and Uehara, T and Ueno, K and Ueshima, G and Uraguchi, F and Ushiba, T and van Putten, M H P M and Vocca, H and Wang, J and Wu, C and Wu, H and Wu, S and Xu, W- R and Yamada, T and Yamamoto, K and Yamamoto, K and Yamamoto, T and Yokogawa, K and Yokoyama, J and Yokozawa, T and Yoshioka, T and Yuzurihara, H and Zeidler, S and Zhao, Y and Zhu, Z -H},
    title = {Overview of KAGRA: Detector design and construction history},
    journal = {Progress of Theoretical and Experimental Physics},
    volume = {2021},
    number = {5},
    pages = {05A101},
    year = {2021},
    month = {05},
    issn = {2050-3911},
    doi = {10.1093/ptep/ptaa125},
    url = {https://doi.org/10.1093/ptep/ptaa125},
    eprint = {https://academic.oup.com/ptep/article-pdf/2021/5/05A101/37974994/ptaa125.pdf},
}

@article{Cho_2018,
   title={Gravitational wave searches for aligned-spin binary neutron stars using nonspinning templates},
   volume={72},
   ISSN={1976-8524},
   url={http://dx.doi.org/10.3938/jkps.72.1},
   DOI={10.3938/jkps.72.1},
   number={1},
   journal={Journal of the Korean Physical Society},
   publisher={Korean Physical Society},
   author={Cho, Hee-Suk and Lee, Chang-Hwan},
   year={2018},
   month=jan, pages={1–5} }

@article{KAGRA:2021vkt,
    author = "Abbott, R. and others",
    collaboration = "KAGRA, VIRGO, LIGO Scientific",
    title = "{GWTC-3: Compact Binary Coalescences Observed by LIGO and Virgo during the Second Part of the Third Observing Run}",
    eprint = "2111.03606",
    archivePrefix = "arXiv",
    primaryClass = "gr-qc",
    reportNumber = "LIGO-P2000318",
    doi = "10.1103/PhysRevX.13.041039",
    journal = "Phys. Rev. X",
    volume = "13",
    number = "4",
    pages = "041039",
    year = "2023"
}

@article{JANKA_2007,
	doi = {10.1016/j.physrep.2007.02.002},
	url = {https://doi.org/10.1016\%2Fj.physrep.2007.02.002},
	year = 2007,
	month = {apr},
	publisher = {Elsevier {BV}},
	volume = {442},
  	number = {1-6},
  	pages = {38--74},
  	author = {H Janka and K Langanke and A Marek and G Martinez Pinedo and B Muller},
  	title = {Theory of core-collapse supernovae},
  	journal = {Physics Reports}
}

@article{Kotake_2006,
	doi = {10.1088/0034-4885/69/4/r03},
  	url = {https://doi.org/10.1088%2F0034-4885%2F69%2F4%2Fr03},
  	year = 2006,
	month = {mar},
  	publisher = {{IOP} Publishing},
 	volume = {69},
 	number = {4},
 	pages = {971--1143},
 	author = {Kei Kotake and Katsuhiko Sato and Keitaro Takahashi},
 	title = {Explosion mechanism, neutrino burst and gravitational wave in core-collapse supernovae},
 	journal = {Reports on Progress in Physics}
}

@article{Kuroda_2016,
	doi = {10.3847/2041-8205/829/1/l14},
 	url = {https://doi.org/10.3847\%2F2041-8205\%2F829\%2F1\%2Fl14},
 	year = 2016,
	month = {sep},
 	publisher = {American Astronomical Society},
 	volume = {829},
 	number = {1},
 	pages = {L14},
 	author = {Takami Kuroda and Kei Kotake and Tomoya Takiwaki},
 	title = {A new Gravitational-Wave signature from standing accretion shock instability in supernovae},
	journal = {The Astrophysical Journal}
}

@article{laura,
    author = "Villegas, Laura O. and Moreno, Claudia and Pajkos, Michael A. and Zanolin, Michele and Antelis, Javier M.",
    title = "{Parameter estimation from the core-bounce phase of rotating core collapse supernovae in real interferometer noise}",
    eprint = "2304.01267",
    archivePrefix = "arXiv",
    primaryClass = "gr-qc",
    doi = "10.1088/1361-6382/add235",
    url={https://iopscience.iop.org/article/10.1088/1361-6382/add235},
    journal = "Class. Quant. Grav.",
    volume = "42",
    number = "11",
    pages = "115001",
    year = "2025"
}

@article{Pajkos_2021,
   title={Determining the Structure of Rotating Massive Stellar Cores with Gravitational Waves},
   volume={914},
   ISSN={1538-4357},
   url={http://dx.doi.org/10.3847/1538-4357/abfb65},
   DOI={10.3847/1538-4357/abfb65},
   number={2},
   journal={The Astrophysical Journal},
   publisher={American Astronomical Society},
   author={Pajkos, Michael A. and Warren, MacKenzie L. and Couch, Sean M. and O’Connor, Evan P. and Pan, Kuo-Chuan},
   year={2021}}

@article{Sharma_2024,
   title={Template bank to search for exotic gravitational wave signals from astrophysical compact binaries},
   volume={109},
   ISSN={2470-0029},
   url={http://dx.doi.org/10.1103/PhysRevD.109.124049},
   DOI={10.1103/physrevd.109.124049},
   number={12},
   journal={Physical Review D},
   publisher={American Physical Society (APS)},
   author={Sharma, Abhishek and Roy, Soumen and Sengupta, Anand S.},
   year={2024},
   month=jun }

@misc{corsi2024multimessengerastrophysicsblackholes,
      title={Multi-messenger Astrophysics of Black Holes and Neutron Stars as Probed by Ground-based Gravitational Wave Detectors: From Present to Future}, 
      author={Alessandra Corsi and Lisa Barsotti and Emanuele Berti and Matthew Evans and Ish Gupta and Konstantinos Kritos and Kevin Kuns and Alexander H. Nitz and Benjamin J. Owen and Binod Rajbhandari and Jocelyn Read and Bangalore S. Sathyaprakash and David H. Shoemaker and Joshua R. Smith and Salvatore Vitale},
      year={2024},
      eprint={2402.13445},
      archivePrefix={arXiv},
      primaryClass={astro-ph.HE},
      url={https://arxiv.org/abs/2402.13445}, 
}

@article{Lee_2015,
    author = "Corsi, Alessandra and others",
    title = "{Multi-messenger astrophysics of black holes and neutron stars as probed by ground-based gravitational wave detectors: from present to future}",
    eprint = "2402.13445",
    archivePrefix = "arXiv",
    primaryClass = "astro-ph.HE",
    doi = "10.3389/fspas.2024.1386748",
    journal = "Front. Astron. Space Sci.",
    volume = "11",
    pages = "1386748",
    year = "2024"
}

@article{shibagaki2020new,
    author = {Shibagaki, Shota and Kuroda, Takami and Kotake, Kei and Takiwaki, Tomoya},
    title = {A new gravitational-wave signature of low-T/W instability in rapidly rotating stellar core collapse},
    journal = {Monthly Notices of the Royal Astronomical Society: Letters},
    volume = {493},
    number = {1},
    pages = {L138-L142},
    year = {2020},
    month = {03},
    issn = {1745-3925},
    doi = {10.1093/mnrasl/slaa021},
    url = {https://doi.org/10.1093/mnrasl/slaa021},
    eprint = {https://academic.oup.com/mnrasl/article-pdf/493/1/L138/56979366/mnrasl_493_1_l138.pdf},
}

@article{Couch_2014,
doi = {10.1088/0004-637X/785/2/123},
url = {https://doi.org/10.1088/0004-637X/785/2/123},
year = {2014},
month = {apr},
publisher = {The American Astronomical Society},
volume = {785},
number = {2},
pages = {123},
author = {Couch, Sean M. and O'Connor, Evan P.},
title = {
High Resolution three-dimensional simulations of core-collase supernovae in multiple progenitors},
journal = {The Astrophysical Journal}
}

@article{Lella_2026,
  title = {Gravitational-wave signals for supernova explosions of three-dimensional progenitors},
  author = {Lella, Alessandro and Lucente, Giuseppe and Kresse, Daniel and Glas, Robert and Janka, Hans-Thomas and Mirizzi, Alessandro},
  journal = {Phys. Rev. D},
  volume = {113},
  issue = {8},
  pages = {083034},
  numpages = {37},
  year = {2026},
  month = {Apr},
  publisher = {American Physical Society},
  doi = {10.1103/f3n4-k2cq},
  url = {https://link.aps.org/doi/10.1103/f3n4-k2cq}
}

@article{Moore_2014,
	doi = {10.1088/0264-9381/32/1/015014},
  	url = {https://doi.org/10.1088/0264-9381/32/1/015014},
  	year = 2014,
	month = {dec},
  	publisher = {{IOP} Publishing},
  	volume = {32},
  	number = {1},
  	pages = {015014},
  	author = {C J Moore and R H Cole and C P L Berry},
  	title = {Gravitational-wave sensitivity curves},
  	journal = {Classical and Quantum Gravity},
}

@article{Gilkis_2017,
   title={Asymmetric core collapse of rapidly rotating massive star},
   volume={474},
   ISSN={1365-2966},
   url={http://dx.doi.org/10.1093/mnras/stx2934},
   DOI={10.1093/mnras/stx2934},
   number={2},
   journal={Monthly Notices of the Royal Astronomical Society},
   publisher={Oxford University Press (OUP)},
   author={Gilkis, Avishai},
   year={2017},
   month=Nov, pages={2419–2429} }

@article{Ott_2012,
   title={Correlated gravitational wave and neutrino signals from general-relativistic rapidly rotating iron core collapse},
   volume={86},
   ISSN={1550-2368},
   url={http://dx.doi.org/10.1103/PhysRevD.86.024026},
   DOI={10.1103/physrevd.86.024026},
   number={2},
   journal={Physical Review D},
   publisher={American Physical Society (APS)},
   author={Ott, C. D. and Abdikamalov, E. and O’Connor, E. and Reisswig, C. and Haas, R. and Kalmus, P. and Drasco, S. and Burrows, A. and Schnetter, E.},
   year={2012}}

@article{Punturo_2010,
doi = {10.1088/0264-9381/27/19/194002},
url = {https://doi.org/10.1088/0264-9381/27/19/194002},
year = {2010},
month = {sep},
publisher = {},
volume = {27},
number = {19},
pages = {194002},
author = {Punturo, M and Abernathy, M and Acernese, F and Allen, B and Andersson, N and Arun, K and Barone, F and Barr, B and Barsuglia, M and Beker, M and Beveridge, N and Birindelli, S and Bose, S and Bosi, L and Braccini, S and Bradaschia, C and Bulik, T and Calloni, E and Cella, G and Mottin, E Chassande and Chelkowski, S and Chincarini, A and Clark, J and Coccia, E and Colacino, C and Colas, J and Cumming, A and Cunningham, L and Cuoco, E and Danilishin, S and Danzmann, K and De Luca, G and De Salvo, R and Dent, T and De Rosa, R and Di Fiore, L and Di Virgilio, A and Doets, M and Fafone, V and Falferi, P and Flaminio, R and Franc, J and Frasconi, F and Freise, A and Fulda, P and Gair, J and Gemme, G and Gennai, A and Giazotto, A and Glampedakis, K and Granata, M and Grote, H and Guidi, G and Hammond, G and Hannam, M and Harms, J and Heinert, D and Hendry, M and Heng, I and Hennes, E and Hild, S and Hough, J and Husa, S and Huttner, S and Jones, G and Khalili, F and Kokeyama, K and Kokkotas, K and Krishnan, B and Lorenzini, M and Lück, H and Majorana, E and Mandel, I and Mandic, V and Martin, I and Michel, C and Minenkov, Y and Morgado, N and Mosca, S and Mours, B and Müller–Ebhardt, H and Murray, P and Nawrodt, R and Nelson, J and Oshaughnessy, R and Ott, C D and Palomba, C and Paoli, A and Parguez, G and Pasqualetti, A and Passaquieti, R and Passuello, D and Pinard, L and Poggiani, R and Popolizio, P and Prato, M and Puppo, P and Rabeling, D and Rapagnani, P and Read, J and Regimbau, T and Rehbein, H and Reid, S and Rezzolla, L and Ricci, F and Richard, F and Rocchi, A and Rowan, S and Rüdiger, A and Sassolas, B and Sathyaprakash, B and Schnabel, R and Schwarz, C and Seidel, P and Sintes, A and Somiya, K and Speirits, F and Strain, K and Strigin, S and Sutton, P and Tarabrin, S and Thüring, A and van den Brand, J and van Leewen, C and van Veggel, M and van den Broeck, C and Vecchio, A and Veitch, J and Vetrano, F and Vicere, A and Vyatchanin, S and Willke, B and Woan, G and Wolfango, P and Yamamoto, K},
title = {The Einstein Telescope: a third-generation gravitational wave observatory},
journal = {Classical and Quantum Gravity}
}

@article{Martynov2016,
  title = {Sensitivity of the Advanced LIGO detectors at the beginning of gravitational wave astronomy},
  author = {Martynov, D. V. and Hall, E. D. and Abbott, B. P. and Abbott, R. and Abbott, T. D. and Adams, C. and Adhikari, R. X. and Anderson, R. A. and Anderson, S. B. and Arai, K. and Arain, M. A. and Aston, S. M. and Austin, L. and Ballmer, S. W. and Barbet, M. and Barker, D. and Barr, B. and Barsotti, L. and Bartlett, J. and Barton, M. A. and Bartos, I. and Batch, J. C. and Bell, A. S. and Belopolski, I. and Bergman, J. and Betzwieser, J. and Billingsley, G. and Birch, J. and Biscans, S. and Biwer, C. and Black, E. and Blair, C. D. and Bogan, C. and Bond, C. and Bork, R. and Bridges, D. O. and Brooks, A. F. and Brown, D. D. and Carbone, L. and Celerier, C. and Ciani, G. and Clara, F. and Cook, D. and Countryman, S. T. and Cowart, M. J. and Coyne, D. C. and Cumming, A. and Cunningham, L. and Damjanic, M. and Dannenberg, R. and Danzmann, K. and Costa, C. F. Da Silva and Daw, E. J. and DeBra, D. and DeRosa, R. T. and DeSalvo, R. and Dooley, K. L. and Doravari, S. and Driggers, J. C. and Dwyer, S. E. and Effler, A. and Etzel, T. and Evans, M. and Evans, T. M. and Factourovich, M. and Fair, H. and Feldbaum, D. and Fisher, R. P. and Foley, S. and Frede, M. and Freise, A. and Fritschel, P. and Frolov, V. V. and Fulda, P. and Fyffe, M. and Galdi, V. and Giaime, J. A. and Giardina, K. D. and Gleason, J. R. and Goetz, R. and Gras, S. and Gray, C. and Greenhalgh, R. J. S. and Grote, H. and Guido, C. J. and Gushwa, K. E. and Gustafson, E. K. and Gustafson, R. and Hammond, G. and Hanks, J. and Hanson, J. and Hardwick, T. and Harry, G. M. and Haughian, K. and Heefner, J. and Heintze, M. C. and Heptonstall, A. W. and Hoak, D. and Hough, J. and Ivanov, A. and Izumi, K. and Jacobson, M. and James, E. and Jones, R. and Kandhasamy, S. and Karki, S. and Kasprzack, M. and Kaufer, S. and Kawabe, K. and Kells, W. and Kijbunchoo, N. and King, E. J. and King, P. J. and Kinzel, D. L. and Kissel, J. S. and Kokeyama, K. and Korth, W. Z. and Kuehn, G. and Kwee, P. and Landry, M. and Lantz, B. and Le Roux, A. and Levine, B. M. and Lewis, J. B. and Lhuillier, V. and Lockerbie, N. A. and Lormand, M. and Lubinski, M. J. and Lundgren, A. P. and MacDonald, T. and MacInnis, M. and Macleod, D. M. and Mageswaran, M. and Mailand, K. and M\'arka, S. and M\'arka, Z. and Markosyan, A. S. and Maros, E. and Martin, I. W. and Martin, R. M. and Marx, J. N. and Mason, K. and Massinger, T. J. and Matichard, F. and Mavalvala, N. and McCarthy, R. and McClelland, D. E. and McCormick, S. and McIntyre, G. and McIver, J. and Merilh, E. L. and Meyer, M. S. and Meyers, P. M. and Miller, J. and Mittleman, R. and Moreno, G. and Mueller, C. L. and Mueller, G. and Mullavey, A. and Munch, J. and Murray, P. G. and Nuttall, L. K. and Oberling, J. and O'Dell, J. and Oppermann, P. and Oram, Richard J. and O'Reilly, B. and Osthelder, C. and Ottaway, D. J. and Overmier, H. and Palamos, J. R. and Paris, H. R. and Parker, W. and Patrick, Z. and Pele, A. and Penn, S. and Phelps, M. and Pickenpack, M. and Pierro, V. and Pinto, I. and Poeld, J. and Principe, M. and Prokhorov, L. and Puncken, O. and Quetschke, V. and Quintero, E. A. and Raab, F. J. and Radkins, H. and Raffai, P. and Ramet, C. R. and Reed, C. M. and Reid, S. and Reitze, D. H. and Robertson, N. A. and Rollins, J. G. and Roma, V. J. and Romie, J. H. and Rowan, S. and Ryan, K. and Sadecki, T. and Sanchez, E. J. and Sandberg, V. and Sannibale, V. and Savage, R. L. and Schofield, R. M. S. and Schultz, B. and Schwinberg, P. and Sellers, D. and Sevigny, A. and Shaddock, D. A. and Shao, Z. and Shapiro, B. and Shawhan, P. and Shoemaker, D. H. and Sigg, D. and Slagmolen, B. J. J. and Smith, J. R. and Smith, M. R. and Smith-Lefebvre, N. D. and Sorazu, B. and Staley, A. and Stein, A. J. and Stochino, A. and Strain, K. A. and Taylor, R. and Thomas, M. and Thomas, P. and Thorne, K. A. and Thrane, E. and Tokmakov, K. V. and Torrie, C. I. and Traylor, G. and Vajente, G. and Valdes, G. and van Veggel, A. A. and Vargas, M. and Vecchio, A. and Veitch, P. J. and Venkateswara, K. and Vo, T. and Vorvick, C. and Waldman, S. J. and Walker, M. and Ward, R. L. and Warner, J. and Weaver, B. and Weiss, R. and Welborn, T. and We\ss{}els, P. and Wilkinson, C. and Willems, P. A. and Williams, L. and Willke, B. and Wilmut, I. and Winkelmann, L. and Wipf, C. C. and Worden, J. and Wu, G. and Yamamoto, H. and Yancey, C. C. and Yu, H. and Zhang, L. and Zucker, M. E. and Zweizig, J.},
  journal = {Phys. Rev. D},
  volume = {93},
  issue = {11},
  pages = {112004},
  numpages = {19},
  year = {2016},
  month = {Jun},
  publisher = {American Physical Society},
  doi = {10.1103/PhysRevD.93.112004},
  url = {https://link.aps.org/doi/10.1103/PhysRevD.93.112004}
}

@ARTICLE{O_Connor_2018,
doi = {10.3847/1538-4357/aaa893},
url = {https://doi.org/10.3847/1538-4357/aaa893},
year = {2018},
month = {feb},
publisher = {The American Astronomical Society},
volume = {854},
number = {1},
pages = {63},
author = {O’Connor, Evan P. and Couch, Sean M.},
title = {Two-dimensional Core-collapse Supernova Explosions Aided by General Relativity with Multidimensional Neutrino Transport},
journal = {The Astrophysical Journal}
}

@article{dimmelmeier2008gravitational,
  title = {Gravitational wave burst signal from core collapse of rotating stars},
  author = {Dimmelmeier, Harald and Ott, Christian D. and Marek, Andreas and Janka, H.-Thomas},
  journal = {Phys. Rev. D},
  volume = {78},
  issue = {6},
  pages = {064056},
  numpages = {28},
  year = {2008},
  month = {Sep},
  publisher = {American Physical Society},
  doi = {10.1103/PhysRevD.78.064056},
  url = {https://link.aps.org/doi/10.1103/PhysRevD.78.064056}
}

@article{Richers_2017,
  title = {Equation of state effects on gravitational waves from rotating core collapse},
  author = {Richers, Sherwood and Ott, Christian D. and Abdikamalov, Ernazar and O'Connor, Evan and Sullivan, Chris},
  journal = {Phys. Rev. D},
  volume = {95},
  issue = {6},
  pages = {063019},
  numpages = {29},
  year = {2017},
  month = {Mar},
  publisher = {American Physical Society},
  doi = {10.1103/PhysRevD.95.063019},
  url = {https://link.aps.org/doi/10.1103/PhysRevD.95.063019}
}

@article{Szczepa_czyk_2021,
	doi = {10.1103/physrevd.104.102002},
  
	url = {https://doi.org/10.1103%2Fphysrevd.104.102002},
  
	year = 2021,
	month = {nov},
  
	publisher = {American Physical Society ({APS})},
  
	volume = {104},
  
	number = {10},
  
	author = {Marek J. Szczepa{\'{n}}czyk and others},
  
	title = {Detecting and reconstructing gravitational waves from the next galactic core-collapse supernova in the advanced detector era},
  
	journal = {Physical Review D}
}

@article{Srivastava_2019,
	doi = {10.1103/physrevd.100.043026},
  	url = {https://doi.org/10.1103%2Fphysrevd.100.043026},
  	year = 2019,
	month = {aug},
  	publisher = {American Physical Society ({APS})},
  	volume = {100},
  	number = {4},
  	author = {Varun Srivastava and Stefan Ballmer and Duncan A. Brown and Chaitanya Afle and Adam Burrows and David Radice and David Vartanyan},
  	title = {Detection prospects of core-collapse supernovae with supernova-optimized third-generation gravitational-wave detectors},
  	journal = {Physical Review D}
}

@article{Vitale_2010,
   title={Parameter estimation from gravitational waves generated by nonspinning binary black holes with laser interferometers: Beyond the Fisher information},
   volume={82},
   ISSN={1550-2368},
   url={http://dx.doi.org/10.1103/PhysRevD.82.124065},
   DOI={10.1103/physrevd.82.124065},
   number={12},
   journal={Physical Review D},
   publisher={American Physical Society (APS)},
   author={Vitale, S. and Zanolin, M.},
   year={2010},
   month={Dec}
}

@article{Zanolin_2010,
   title={Application of asymptotic expansions for maximum likelihood estimators errors to gravitational waves from binary mergers: The single interferometer case},
   volume={81},
   ISSN={1550-2368},
   url={http://dx.doi.org/10.1103/PhysRevD.81.124048},
   DOI={10.1103/physrevd.81.124048},
   number={12},
   journal={Physical Review D},
   publisher={American Physical Society (APS)},
   author={Zanolin, M. and Vitale, S. and Makris, N.},
   year={2010},
   month={Jun}
}

@techreport{T2000012,
    author = "Abbott, R. and others",
    collaboration = "LIGO Scientific, Virgo",
    title = "{Noise curves used for Simulations in the update of the Observing Scenarios Paper}",
  journal = {LIGO Document Control Center},
     year = 2020,
    month = Feb,
   number = {LIGO-T2000012},
      url = {https://dcc.ligo.org/LIGO-T2000012/public}
}

\end{document}